\documentclass[onecolumn,superscriptaddress,aps]{revtex4}
\usepackage{graphicx}
\usepackage{latexsym}
\usepackage{amsmath}

\begin{document}

\title{Swarms with canonical active Brownian motion}

\author{Alexander Gl\"uck} \email{alexander.glueck@chello.at}
\affiliation{Faculty of Physics, University of Vienna,
   Boltzmanngasse 5, A-1090 Wien}
\author{Helmuth H\"uffel} % \email{helmuth.hueffel@univie.ac.at}
\affiliation{Faculty of Physics, University of Vienna,
   Boltzmanngasse 5, A-1090 Wien}
\author{Sa{\v s}a Iliji{\'c}} % \email{sasa.ilijic@fer.hr}
\affiliation{Department of Physics,
   Faculty of Electrical Engineering and Computing,
   University of Zagreb, Unska 3, HR-10000 Zagreb}
% Two of us, HH and SI, wish to comment-out our email addresses
% not to be burried with the terrible dagger symbols
% forced by the default behaviour of RevTeX4.

\begin{abstract}
We present a swarm model of Brownian particles with harmonic interactions, where the individuals undergo canonical active Brownian motion, i.e. each Brownian particle can convert internal energy to mechanical energy of motion. We assume the existence of a single global internal energy of the system. Numerical simulations show amorphous swarming behavior as well as static configurations. Analytic understanding of the system is provided by studying stability properties of equilibria.
\end{abstract}

\maketitle

\section{Canonical active Brownian motion}

The concept of active Brownian motion was developed to describe complex motion in various systems \cite{schw und ebe}, \cite{schw}. Its main idea is to assume the existence of an additional \textit{internal} degree of freedom, called the 'internal energy' $e$, which can be taken up from the environment and converted to energy of motion. Former models only assumed an exchange between the internal energy and the \textit{kinetic} energy of motion. In \cite{paper}, we discussed a more general case, where $e$ can be also converted to the potential energy of the particle, thus to the \textit{full} mechanical energy $H=\frac{p^2}{2}+U(x)$. The corresponding stochastic equations for the position $\mathbf{x} \in \textbf{R}^3$ and the momentum $\mathbf{p}$ of the particle read
\begin{equation}
\frac{d\mathbf{x}}{dt}=\frac{\partial H}{\partial \mathbf{p}}, \quad \frac{d\mathbf{p}}{dt}=-\left(  1-d_1 e \right) \frac{\partial H}{\partial \mathbf{x}} - \left(  \gamma - d_2 e \right) \frac{\partial H}{\partial \mathbf{p}} + \mbox{\boldmath $\eta$},
\end{equation}
\begin{equation}
\frac{de}{dt}=1 - c e - e H,
\end{equation}
where $\mbox{\boldmath $\eta$}$ represents the stochastic noise, satisfying the correlations
\begin{equation}
\langle \eta ^a \rangle = 0, \quad \langle \eta^a (t) \eta^b (\bar{t}) \rangle = 2 \delta^{ab} \delta(t-\bar{t})
\end{equation}
and therefore describing a completely irregular force. The constants $d_i$ and $c$ remain to be fixed for concrete applications. \\ \\The above system of equations defines the dynamics of a particle subject to an external potential $U(x)$ and an irregular force $\mbox{\boldmath $\eta$}$. The friction with the environment is characterized by the friction constant $\gamma$. The dynamical equation for $e$ has the following meaning: The first term gives the constant rate of internal energy take-up from the environment, the second term $-c e$ models the inner energy loss of the indivual, and the third term $- e H$describes the conversion of internal energy $e$ into the mechanical energy $H$. Numerical simulations and analytic investigations of these dynamics were made in \cite{paper} for a single particle in harmonic external potentials. Studies of the multi-particle case were missing so far. \\ \\ In \cite{swarm} many particle systems with active Brownian motion were discussed and a specific form of circular swarming behavior was found. In this case, however, internal energy was converted only to \textit{kinetic} energy (corresponding to $d_1=0$ and $H$ being replaced by $\frac{p^2}{2}$ in the equation for $\dot{e}$). Further simplifications were made by studying only situations, where the internal energy $e$ was considered to be stationary ($\frac{de}{dt}=0$) and thus could be eliminated from the dynamical equations. \\ \\We now want to discuss in full generality the behavior of many particles undergoing \textit{canonical} active Brownian motion, i.e. study particles which can convert dynamical internal energy to \textit{total} mechanical energy. We will show amorphous swarming behavior as well as static configurations.

\section{Many particle systems}

We present a simple model for a swarm of active Brownian particles by assuming the existence of a single global internal energy $e$ of the swarm. Consider the movement of $n$ active Brownian particles, enumerated by the index $i$, with unit mass in $d$ space dimensions coupled to the internal energy $e$. The position vectors are $\textbf{x}_i \in \textbf{R}^d$ with $i=1,\dots,n$ and the stochastic equations of motion read
\begin{equation}
\frac{d\textbf{x}_i}{dt}=\frac{\partial H}{\partial \textbf{p}_i},
\end{equation}
\begin{equation}
\frac{d\textbf{p}_i}{dt}=-\left(  1 - d_1 e  \right) \frac{\partial H}{\partial \textbf{x}_i} - \left(  \gamma - d_2  e \right) \frac{\partial H}{\partial \textbf{p}_i} + \mbox{\boldmath $\eta$}_i,
\end{equation}
\begin{equation}
\frac{de}{dt}=1-c e - e H,
\end{equation}
where the total mechanical energy is given by
\begin{equation}
H=\sum_{i=1}^{n} \left( \frac{ \textbf{p}_i ^2}{2} + U_i^{(in)} + U_i^{(ex)} \right).
\end{equation}
The potential energy of one particle is composed of an potential $U_i^{(in)} $, describing the interaction with all the other particles, and an external potential $U_i^{(ex)}$, modeling the environment of the swarm. We study the case where all individuals are attracted to the center of mass $\textbf{X}= \frac{1}{n} \sum_{i=1}^n \textbf{x}_i$ and are moving in a landscape of harmonic shape. The total mechanical energy therefore reads
\begin{equation}
H = \sum_{i=1}^{n} \left(  \frac{ \textbf{p}_i ^2}{2} + \sum_{j=1} ^{n} \frac{k_1}{4n} \left(  \textbf{x}_i - \textbf{x}_j \right)^2 + \frac{k_2}{2} \textbf{x}_i ^2  \right),
\end{equation}
(with positive constants $k_1$ and $k_2$) so that the swarm dynamics is finally given by the equations
\begin{eqnarray}
\dot{\textbf{x}_i} & = & \textbf{p}_i, \label{ui} \\
\dot{\textbf{p}_i} & = & -\left(  1 - d_1 e  \right) \left( k_1 \left( \textbf{x}_i - \textbf{X} \right) + k_2 \textbf{x}_i \right) - \left(  \gamma - d_2  e \right) \textbf{p}_i +  \mbox{\boldmath $\eta$}_i, \\
\dot{e} & = & 1-c e - e \sum_{i=1}^{n} \left(  \frac{ \textbf{p}_i ^2}{2} + \sum_{j=1} ^{n} \frac{k_1}{4n} \left(  \textbf{x}_i - \textbf{x}_j \right)^2 + \frac{k_2}{2} \textbf{x}_i ^2  \right). \label{oi}
\end{eqnarray}

\subsection{Variable transformation}

The above system of coupled stochastic nonlinear differential equations is not easily accessible with direct analytic procedures. Nonetheless, predictions for the long time behavior can be made by transforming to the new variables
\begin{equation}
K=\sum_{i=1}^{n} \frac{\textbf{p}_i ^2}{2}, \quad U = \sum_{i=1}^{n} \frac{\textbf{x}_i ^2}{2}, \quad S = \sum_{i=1}^{n} \textbf{x}_i \textbf{p}_i,
\end{equation}
\begin{equation}
L = \left(  \sum_{i=1}^n \mathbf{p}_i \right)^2, \quad V = \left(  \sum_{i=1}^n \mathbf{x}_i \right)^2, \quad T = \left( \sum_{i=1}^n \textbf{x}_i \right) \left( \sum_{i=1}^n \textbf{p}_i \right).
\end{equation}
$K$ represents the total kinetic energy of the swarm, $U$ the total external potential energy, whereas the variables $V$ and $L$ are quadratic forms of the center of mass $\textbf{X}$, resp. its time derivative $\textbf{P}\equiv\frac{d\textbf{X}}{dt}$: $V=n^2 \textbf{X}^2$ and $L=n^2 \textbf{P}^2$. Additionally, also the variable $T$ can be written in terms of $\textbf{X}$ and $\textbf{P}$: $T=n^2 \textbf{X} \textbf{P}$. In these new variables, when omitting the noise influences, the system reads:
\begin{eqnarray}
\dot{K} & = & -  \left(  1-d_1 e  \right) \left(   \left(  k_1+k_2  \right) S -  \frac{k_1}{n} T \right)  - 2 \left(  \gamma - d_2 e \right) K, \label{eins} \\
\dot{U} & = & S, \\
\dot{S} & = & 2 K - \left( 1-d_1 e  \right)  \left(  2 \left(  k_1 + k_2 \right) U -  \frac{k_1}{n} V \right) - \left(  \gamma - d_2 e \right) S, \\
\dot{e} & = & 1 - c e - e \left(  K + \left(  k_1+k_2 \right) U - k_1 \frac{V}{2n}   \right),  \\
\dot{T} & = & L - \left(  1 - d_1 e \right) k_2 V - \left( \gamma - d_2 e \right) T, \\
\dot{V} & = & 2 T, \\
\dot{L} & = & - 2 \left( 1 - d_1 e \right) k_2 T - 2 \left( \gamma - d_2 e \right) L. \label{zwei}
\end{eqnarray}

\subsection{Equilibrium solutions}
To handle the above nonlinear system, we first calculate stationary points, i.e. values of $z=(K,U,S,e,T,V,L)$, where $\dot{z}=0$  and linearize the equations around these equilibrium solutions. By calculating eigenvalues of the Jacobian, we can analyze stability properties. We find three distinct equilibrium points $E_i$:
\begin{eqnarray}
&E_1:& e_0 = \frac{1}{c}, \quad K_0 = U_0 = S_0 = e_0 = T_0 = V_0 = L_0 = 0. \nonumber \\ \nonumber
&E_2:& U_0 = \frac{ d_1 - c + \frac{k_1 V_0}{2n}  }{ ( k_1 + k_2)}, \quad e_0 = \frac{1}{d_1}, \quad K_0 = S_0 = T_0 = L_0 = 0. \\
&E_3:& U_0 = \frac{  d_2 \left( d_2 - c \gamma \right) + \frac{\gamma k_1}{ 2 n } \left( 2 d_2 - \gamma d_1 \right) V_0}{\gamma ( k_1 + k_2) \left( 2 d_2 - \gamma d_1 \right) }, \quad e_0 = \frac{\gamma}{d_2}, \nonumber \\ \nonumber
&   & L_0 = k_2 \left( 1- \gamma \frac{d_1}{d_2} \right) V_0, \quad S_0 = T_0 = 0, \nonumber \\ \nonumber
&   & K_0 = \frac{(\gamma d_1 - d_2)(\gamma c - d_2)}{\gamma (2d_2 - \gamma d_1)}.
\end{eqnarray}
Equilibria $E_1$ and $E_2$ are truly \textit{static} configurations of the swarm, since $K_0=0$ in both cases, whereas equilibirum $E_3$ has a nonvanishing equilibrium value of $K_0$, which means that swarm particles are actually moving. \\ \\ Calculating the stability conditions of all three equilibria allows us to decide under which parameter conditions the swarm either collapses to a point ($E_1$), freezes to a certain pattern ($E_2$), or moves amorphously ($E_3$). For equilibrium $E_1$, all seven eigenvalues of the Jacobian are accessible and it is found to be stable under the conditions
\begin{equation}
\gamma > \frac{d_2}{c}, \quad c > d_1, \quad 4 c \left(  c- d_1 \right)  \left( k_1 + k_2 \right)    \leq \left(  \gamma c - d_2 \right)^2.
\end{equation}
If these inequalities are satisfied, all swarm particles will collapse into the origin and will remain motionless apart from stochastic fluctuations.
\begin{figure}
\begin{center}
\includegraphics[scale=1.1]{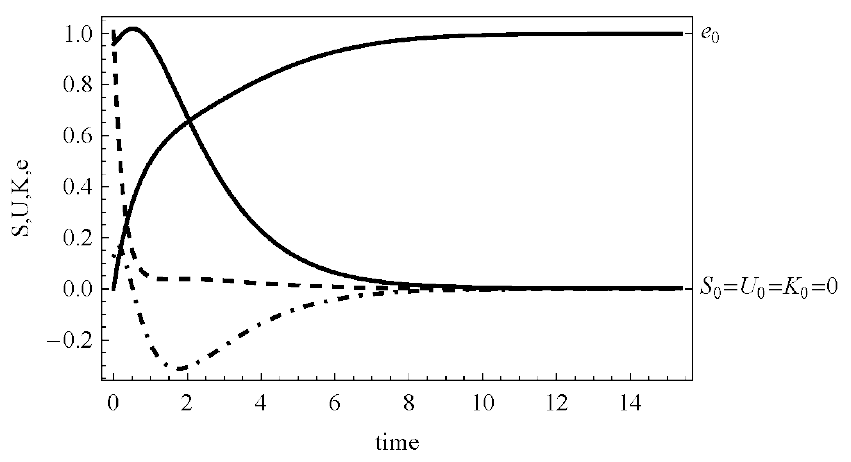}
\end{center}
\caption{\label{fig: one} Time evolution of the deterministic equations (\ref{eins})-(\ref{zwei}) for $K$ (dashed line), $U$ (solid-line), $S$ (dot-dashed line), $e$ (solid line), $T$, $V$ and $L$ (not shown for simplicity of presentation) for the collapse of the swarm to the minimum of the external potential ($E_1$). Simulation parameters are $c=1$, $d_1=1/2$, $d_2=1$, $\gamma = 2$, $k_1 = 1/4$ and $k_2=1/4$ ($V_0=0$).}
\end{figure}
The evolution of such a collapse, according to equations (\ref{eins})-(\ref{zwei}) is shown in figure \ref{fig: one}. More interesting is equilibrium $E_2$, which is stable under the conditions $\gamma > \frac{d_2}{d_1}$, $c < d_1$ and
\begin{equation}
2 \left(  d_1 - c \right) \left( k_1 + k_2 \right) < \left(  d_1 \gamma - d_2   \right) \left(  d_1 + \gamma - \frac{d_2}{d_1} \right).
\end{equation}
We used the Routh-Hurwitz-Theorem \cite{routh} to determine the stability properties. Exactly one eigenvalue is identical to zero independet of the parameter choice. $E_2$ is in fact not an isolated fixpoint, but a one-dimensional continuum of equilibria, since $U_0$ and $V_0$ are linearly dependent. This corresponds to exactly one zero-eigenvalue of the Jacobian, whereas the other six eigenvalues are stable under the conditions given above. According to stable manifold theory, for every chosen value of $V_0$, there exists a stable manifold on which the trajectory converges asymptotically to equilibrium $E_2$. (In the following, we have set $V_0=0$.) In this second equilibrium state, the whole swarm is frozen to a static configuration, which satisfies $\sum_{i=1}^{n} \textbf{x}_i ^2 =$ const.
\begin{figure}
\begin{center}
\includegraphics[scale=1.1]{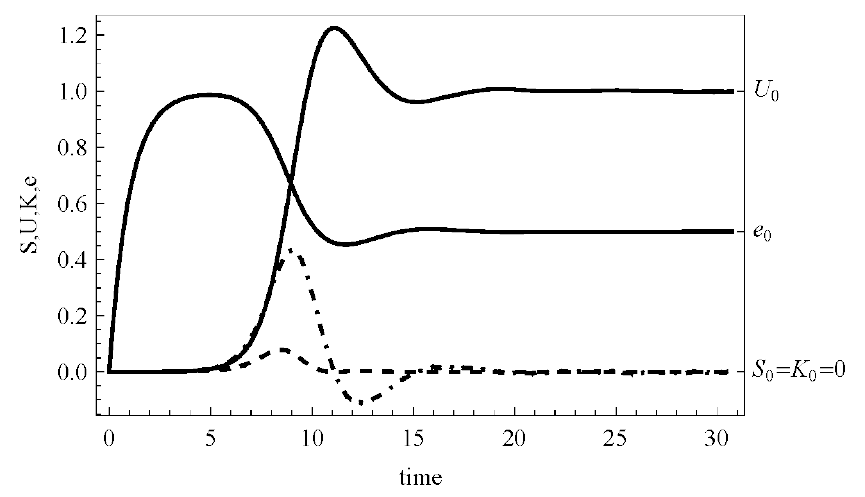}
\end{center}
\caption{\label{fig: two} Relaxation to a jelly-like state ($E_2$), according to equations (\ref{eins})-(\ref{zwei}). Simulation parameters are $c=1$, $d_1=2$, $d_2=1$, $\gamma = 2$, $k_1 = 1/4$ and $k_2=1/4$ ($V_0=0$).}
\end{figure}
One example of such a frozen state is shown in figure \ref{fig: two}. If the swarm is subject to noise influences, the individual particles will oscillate around their fixed positions. \\ \\ \textit{Dynamic} states of the swarm are realized when reaching equilibrium $E_3$, since the sum of all kinetic energies then has a non-vanishing value. $E_3$ is found to be stable under the conditions $d_1 < 0$, $\gamma < \frac{d_1}{c}$ and
\begin{equation}
\gamma d_1 \left( k_1 + k_2\right) < \frac{ d_2^2\left(  d_2 - d_1 \gamma \right) \left( 2 d_2^2 - 2d_1 d_2 \gamma + c d_1 \gamma  \right)}{\left(  4 d_2^3 - 6 \gamma d_1 d_2^2 + 3 \gamma ^2 d_1^2 d_2 - c \gamma ^3 d_1^2   \right)}.
\end{equation}
Again, one zero eigenvalue appears and the Ruth-Hurwitz-Theorem was applied to determine the stability of the system. Note that in $E_3$, the two equations for $U_0$ and $L_0$ define two planes in the space of variables $(U,V,L)$, whose intersection gives a line, i.e. a manifold of equilibria with dimension one, which matches the overall number of zero eigenvalues. Hence the theory of stable manifolds again ensures that we have converging trajectories to $E_3$ under the conditions given above. If these inequalities are satisfied by a specific parameter choice, the swarm reaches equilibrium $E_3$ and appears to simulate the amorphous behavior of insect swarms.
\begin{figure}
\begin{center}
\includegraphics[scale=1.1]{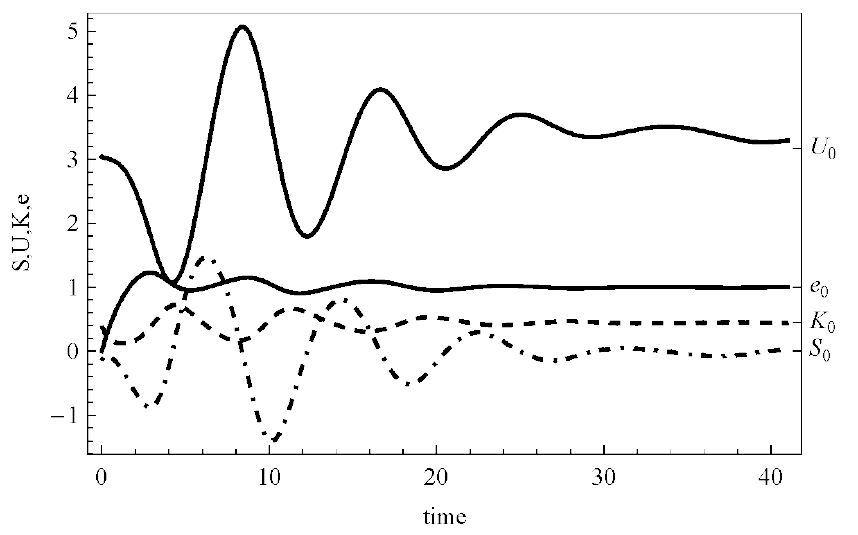}
\end{center}
\caption{\label{fig: three} State of amorphous swarming ($E_3$), according to equations (\ref{eins})-(\ref{zwei}). Simulation parameters are $c=1/3$, $d_1=-1$, $d_2=1$, $\gamma = 1$, $k_1 = 1/16$ and $k_2=1/128$ ($V_0=0$).}
\end{figure}
In figure \ref{fig: three}, the equilibrating process is explicitly shown. \\ \\ Numerical simulation of our original equations (\ref{ui})-(\ref{oi}) gives the full $N$-particle time evolution. Video examples for the corresponding motion of swarm particles are available online \cite{web} for all three distinct cases $E_1$ (collapse), $E_2$ (jelly-like-state) and $E_3$ (amorphous swarming). Simulations of the stochastic system (\ref{ui})-(\ref{oi}) in two dimensions are shown, which - apart of small noise perturbations - agree nicely with our simulations of the deterministic equations (\ref{eins})-(\ref{zwei}).

\section{Summary and Outlook}
We have formulated swarm dynamics based on canonical active Brownian motion. We assumed the existence of a single global internal energy of the swarm and postulated its coupling to the full Hamiltonian of the system. We studied the case where all particles were harmonically attracted to their center of mass and were all moving in a landscape of harmonic shape. The system was simulated numerically and analytic understanding was obtained by studying stability properties of equilibria. We found three distinct forms of the large time behavior of the swarm, showing collapse to a point, freezing to a static configuration and amorphous swarming behavior, respectively. \\ \\
An interesting application of our present investigation appears to be the study of particle swarms in given (genuinely complicated) potentials and the development of a new algorithm for particle swarm optimization theory. Furthermore we mention the modeling of crystals and applications in solid state physics.
\\ \\
\textbf{Acknowledgments}: We thank Josef Hofbauer for his valuable advice on dynamical systems. In addition, we are grateful for financial support within the Agreement on Cooperation between the Universities of Vienna and Zagreb.

\end{document}